\documentclass[12pt]{article}
\usepackage[table]{xcolor}
\usepackage{amsmath,amssymb,exscale,cancel}
\usepackage{graphicx}
\usepackage{epsfig}
\usepackage{multicol}
\usepackage{color}
\usepackage{mathrsfs}
\usepackage{blindtext}
 \usepackage{fancyhdr}
\usepackage{hyperref}
\usepackage{cite}
\usepackage{mathtools}

\usepackage{floatrow}

\usepackage{graphicx}
\usepackage{sidecap}

\usepackage[latin1]{inputenc} 
\usepackage{multicol}  
 
 \usepackage{titlesec}
 
 \usepackage{rotating,slashed,xcolor,amsfonts,expdlist,charter}

\numberwithin{equation}{section}

\usepackage{xcolor}
\usepackage{sectsty}

\usepackage{mdframed}
\usepackage{titletoc}

\definecolor{secnum}{RGB}{13,151,225}
\definecolor{ptcbackground}{RGB}{212,237,252}
\definecolor{ptctitle}{RGB}{0,177,235}

\titlecontents{lsection}
  [5.8em]{\sffamily}
  {\color{secnum}\contentslabel{2.3em}\normalcolor}{}
  {\titlerule*[1000pc]{.}\contentspage\\\hspace*{-5.8em}\vspace*{5pt}%
    \color{white}\rule{\dimexpr\textwidth-15.5pt\relax}{1pt}}

\usepackage{hyperref}
\hypersetup{colorlinks,bookmarksopen,bookmarksnumbered,citecolor=rossos,
linkcolor=redy,pdfstartview=FitH,urlcolor=blus}
\usepackage{slashed}

\definecolor{blus}{cmyk}{1,1,0,0.1}
\definecolor{verdes}{cmyk}{0.99,0,0.59,0.65}
\definecolor{rossos}{cmyk}{0,1,1,0.55}
\definecolor{redy}{cmyk}{0,1,1,0.7}
\definecolor{greeny}{cmyk}{0.99,0,0.59,0.98}
\definecolor{green-go}{cmyk}{0.79,0,0.59,0.5}

\usepackage{titlesec}

\newcommand{\beq}{\begin{equation}}
\newcommand{\eeq}{\end{equation}}

\def\hhref#1{\href{http://arxiv.org/abs/#1}{arXiv:#1}} 

\newcommand{\tmtextbf}[1]{{\bfseries{#1}}}
\newcommand{\tmtextrm}[1]{{\rmfamily{#1}}}

\def\be{\begin{equation}}
\def\ee{\end{equation}}
\def\ba{\begin{array} }
\def\bac{\begin{array} {c}}
\def\bacc{\begin{array} {cc}}
\def\baccc{\begin{array} {ccc}}
\def\bacccc{\begin{array} {cccc}}
\def\ea{\end{array}}
\def\bea{\begin{eqnarray}}
\def\eea{\end{eqnarray}}

\definecolor{red}{rgb}{1,0,0}

\def\psl{\hbox{\hbox{${p}$}}\kern-1.9mm{\hbox{${/}$}}}
\def\dsl{\hbox{\hbox{${\partial}$}}\kern-2.2mm{\hbox{${/}$}}}
\def\Dsl{\hbox{\hbox{${D}$}}\kern-2.6mm{\hbox{${/}$}}}

\newcommand{\gappeq}{{\rlap{{\raise}.5ex\text{\ensuremath{>}}}{{\lower}.5ex\text{\ensuremath{\sim}}}}}
\newcommand{\lappeq}{{\rlap{{\raise}.5ex\text{\ensuremath{<}}}{{\lower}.5ex\text{\ensuremath{\sim}}}}}

\newcommand{\I}{\tmtextrm{1{\kern}-.24em l}}

\begin{document}
\topmargin -1.0cm
\oddsidemargin 0.9cm
\evensidemargin -0.5cm

{\vspace{-1cm}}
\begin{center}

\vspace{-1cm}


 {\Huge \tmtextbf{ \color{blus} 
A Fundamental QCD Axion Model}} {\vspace{.5cm}}\\
 
\vspace{0.9cm}

{\large  {\bf Alberto Salvio }
\vspace{.3cm}

{\em  

\vspace{.4cm}
Physics Department, University of Rome and INFN Tor Vergata, Italy\\ 

\vspace{0.4cm}

\vspace{0.2cm}

 \vspace{0.9cm}
}}

\noindent ------------------------------------------------------------------------------------------------------------------------------
 \vspace{-0.7cm}
\end{center}

\begin{center}
{\bf \large Abstract}
\end{center}

  \vspace{0.1cm}
\noindent   We construct and study a fundamental field theory of the QCD axion: all couplings flow to zero in the infinite-energy limit realizing the totally asymptotically free (TAF) scenario. Some observable quantities (such as the masses of new quarks and scalars) are predicted at low energies by the TAF requirement  in terms of gauge couplings and a vector-boson mass. Here the minimal model of this sort is explored; the axion sector is charged under an SU(2) gauge group and a dark photon appears at low energies. This model can be TAF and feature an absolutely stable vacuum at the same time.

\noindent ------------------------------------------------------------------------------------------------------------------------------
%


\vspace{0.cm}
%
\noindent 


\vspace{-.5cm}

\noindent 


\vspace{0.9cm}

\section{Introduction}

\noindent QCD is perhaps the most satisfying building block of the Standard Model (SM). Not only it provides us with an accurate description of strong interactions, but is also a non-trivial fundamental theory: asymptotic freedom~\cite{Gross} tells us  that QCD remains interacting  in the continuum limit.

It is surprising that, while the Yukawa interactions of the SM violate CP, the QCD Lagrangian respects it: the {\it possible} CP-violating  $\theta$ angle, which includes the effect of the phases of the quark mass matrix,
is strongly constrained by the experiments (for a recent review see~\cite{DiLuzio:2020wdo}).  

A possible explanation was proposed by Peccei and Quinn (PQ)~\cite{Peccei:1977hh}: they introduced a global chiral U(1) symmetry (called PQ symmetry and denoted here U(1)$_{\rm PQ}$), under which some  colored particles transform. These can be the quarks of the SM and/or some extra (still unobserved) quarks.  Since all quarks must be massive, U(1)$_{\rm PQ}$ has to be spontaneously broken. The corresponding Goldstone boson~\cite{Weinberg:1977ma}, called the axion, is a good dark matter candidate~\cite{Preskill:1982cy}. 
Being U(1)$_{\rm PQ}$ anomalous, the axion acquires a non-vanishing potential (becoming a pseudo-Goldstone boson) and   the QCD sector lies on a CP-symmetric vacuum.

In order to realize such a breaking in concrete models and preserve computability at the same time  one typically introduces new scalars and thus new quartic couplings (see e.g.~\cite{Kim:1979if,DFSZ}).
However, mainly because of the difficulty in having asymptotically free (AF) quartic couplings, all these field-theoretic axion models proposed so far suffer from a Landau pole (LP) and spoil the asymptotic freedom of QCD. 

The purpose of this paper is to construct and study  a fundamental and realistic field theory of axions, whose  predictions can be computed explicitly from infinite energy down to the QCD confinement scale, below which lattice QCD technology needs to be used anyway. To the best of our knowledge, no previous constructions had all these features {\it at the same time}. For example, there are viable axion models whose perturbative renormalization group equations (RGEs) have LPs, though in some constructions the LPs are all above the Planck scale (see e.g.~\cite{Salvio:2015cja}). It is possible that the LPs are only an artefact of perturbation theory; however, a confirmation of this would require  non-perturbative calculations (e.g.~on the lattice) and currently there is no lattice  evidence that the LPs can disappear in the exact renormalization group flow, at least for axion models. On the other hand, composite axion constructions
such as~\cite{Kim:1984pt} can be viable TAF
models, but they conjecture that  the  vacuum expectation values of the fields respect the required symmetry breaking and evidence
 of the validity of this ansatz (which again would involve non-perturbative calculations) is still missing to the best of our knowledge.  We focus here on a {\it minimal} and computable realistic model that implements U(1)$_{\rm PQ}$ and its breaking (leaving SU(3)$_c$ 
unbroken) and is TAF 
at the same time.  

The TAF requirement has been used in the literature to obtain UV-complete extensions of particle physics models~\cite{K5, Giudice:2014tma,Holdom:2014hla,Pelaggi:2015kna}. One common feature of these constructions is the presence of several extra fields and potentially   further sources of CP violation, which, in the absence of U(1)$_{\rm PQ}$, may induce a too large radiative contribution to $\theta$ unless a  tremendous fine tuning is performed. This is another independent motivation to construct a TAF axion model. Yet another motivation is the fact that TAF models can predict the low-energy values of some observables: this can happen because some couplings must have precise low energy values in order for all couplings to be AF.

In the present paper the other two fine-tuning problems that affect  the SM (the cosmological constant and the Higgs mass ones) are not discussed. The main motivation for doing so is that, while these problems can be addressed with anthropic arguments~\cite{Weinberg:1987dv}, there appears to be no anthropic solution for the strong-CP problem. 

Here we assume that gravitational interactions, unlike what happens in Einstein gravity, become so weak at high energy that their impact on the renormalization group (RG) flow is negligible, but  still all successes of Einstein's theory at accessible energies are  reproduced. In particular, it is assumed that the gravitational couplings (analogous to the gauge couplings in Yang-Mills theories) approach zero before the matter couplings in the UV. This scenario, called softened gravity~\cite{Giudice:2014tma} may be realized, for example, in UV modifications of gravity featuring quadratic curvature terms in the action~\cite{Salvio:2014soa} or in non-local extensions of general relativity~\cite{Frolov:2015bta}. We, therefore, neglect gravity in the present study.

\section{Building the model}

As well-known, the  scalars of a TAF model should be charged under some gauge interaction and the gauge group must not have any U(1) factor to avoid LPs.

The minimal possibility (which we consider here) is having the axion sector gauge invariant under an SU(2) group (henceforth SU(2)$_a$). Then the full gauge group contains the  factor SU(3)$_c\times$SU(2)$_a$, where SU(3)$_c$ is the ordinary SU(3) of strong interactions.  
The gauge group should also include extra factors to account for a TAF extension of the  SM   (explicit realizations were provided in~\cite{Giudice:2014tma,Holdom:2014hla,Pelaggi:2015kna}). We will refer to such an extension as the SM sector. This sector has to be present, in addition to  the axion sector we describe here, for obvious phenomenological reasons. The SM and axion sectors talk to each other through the SU(3)$_c$ gauge interactions.
Here we will take as typical example of TAF SM extensions
those based on the trinification gauge group SU(3)$_L\times$SU(3)$_c\times$SU(3)$_R$~\cite{Pelaggi:2015kna} 
because SU(3)$_c$ is not embedded in a larger gauge group factor, in contrast to other known TAF models like, for instance, those based on the Pati-Salam group  SU(2)$_L\times$SU(4)$_{\rm PS}\times$SU(2)$_R$~\cite{Giudice:2014tma,Holdom:2014hla}. However, we will not commit ourselves to any specific TAF SM extension here.
 
In order to have U(1)$_{\rm PQ}$ invariance we introduce two extra Weyl fermions $q$ and $\bar q$ in the fundamental and antifundamental of  SU(3)$_c\times$SU(2)$_a$,  respectively, and give them  the same PQ charge: $\{q,\bar q\}\rightarrow e^{i\alpha/2}\{q,\bar q\}$, where $\alpha$ is a constant. For the sake of minimality we require the PQ charges of all particles in the SM sector to vanish; from this point of view the model we are constructing is similar to the KSVZ-like axion models~\cite{Kim:1979if}. Since the extra-fermion representation  of the gauge group   is vector like, there are no gauge anomalies as long as the SM sector is free from gauge anomalies; this is clearly the case for the trinification SM sectors, whose fermions can form a representation of the anomaly free $E_6$  group containing SU(3)$_L\times$SU(3)$_c\times$SU(3)$_R$. As usual U(1)$_{\rm PQ}$ forbids an explicit mass term $\bar qq$ and so, in order to give mass to these extra quarks (as required by the experiments), we introduce a scalar field $A$, which spontaneously breaks U(1)$_{\rm PQ}$. Therefore, $A$ has to be complex and have Yukawa interactions with $q$ and $\bar q$,
\be  \mathscr{L}_y = - y \bar qA q +{\rm H.c.}\, . \label{Yukawa} \ee
The PQ symmetry of $\mathscr{L}_y$ requires $A$ to transform  under U(1)$_{\rm PQ}$  as follows: $A\rightarrow e^{-i\alpha}A$. Gauge invariance, on the other hand, tells us that $A$ has to be invariant under SU(3)$_c$ and belong to the adjoint of SU(2)$_a$. The scalar $A$, being complex,  contains two Hermitian adjoint representations $A_R$ and $A_I$ and we can decompose $A= A_R+i A_I$. Note that further Yukawa interactions besides~(\ref{Yukawa}) and those present  in the SM sector are forbidden by the gauge symmetries and ${\rm U(1)}_{\rm PQ}$.

The potential of $A$ is given by
\be V_A= -m^2{\rm Tr}(A^\dagger A)+\lambda_1 {\rm Tr}^2(A^\dagger A)  +
 \lambda_2|{\rm Tr}(A A)|^2,
\label{UA}\ee
where $m^2$ is taken to be positive to trigger the spontaneous breaking of U(1)$_{\rm PQ}$.
Both ${\rm Tr}(A^\dagger A)$ and $|{\rm Tr}(A A)|^2$  are real and non-negative. Therefore, the couplings  $\lambda_i$ (with $i=1,2$)   are real and  vacuum stability at high-field values (henceforth ``high-field stability") is guaranteed for $\lambda_i> 0$. However, these conditions are sufficient but not necessary for high-field stability. Indeed,   since ${\rm Tr}^2(A^\dagger A)\geq|{\rm Tr}(A A)|^2$ the coupling $\lambda_2$ can be negative and the necessary {\it and} sufficient conditions for high-field stability are\footnote{If one extends the extra gauge group beyond SU(2)$_a$ the potential  generically involves more quartic couplings and the stability analysis becomes more challenging.}
\be \lambda_1>0, \qquad \lambda_1+ \lambda_2>0.\label{VacS12}\ee
Later on we will show that this model is TAF and stable at high fields for some values of the parameters  and for the same values absolute vacuum stability (not only high-field stability) is guaranteed.
Here we neglect the couplings with the scalars of the SM sector; we note that setting to zero those couplings  is consistent at the one-loop level because they are not generated and so they remain zero at the one-loop level if their initial conditions in the RG flow is set to zero. The one-loop approximation, on the other hand, is enough for our purposes because total asymptotic freedom implies that all couplings approach zero at high energies. So, in order to establish total asymptotic freedom we can focus on the one-loop RGEs.

%

\section{The RG flow}

The one-loop $\beta$-function of the gauge coupling $g$  of a generic gauge group $G$ is
\be \frac{dg^2}{dt} = -bg^4, \qquad b\equiv\frac{11}{3} C_2(G)-\frac43 S_2(F) -\frac16 	S_2(S),\label{betag}\ee 
where $t\equiv \ln(\mu^2/\mu_0^2)/(4\pi)^2$, the energy scale $\mu_0$ is arbitrary, $\mu$ is the usual RG scale and $C_2(G)$, $S_2(F)$ and $S_2(S)$ are the Dynkin indices of the adjoint representation ($C_2(G)=N$ for $G= $ SU($N$)), the Dirac-spinor representation and the scalar representation, respectively.
The general solution to Eq.~(\ref{betag}) is $g^2(t) = g_0^2/(1+g_0^2 b t)$,
where $g_0\equiv g(0)$. Then in order to have an AF gauge coupling and avoid a LP we must have $b>0$. The corresponding Gaussian fixed point is UV attractive: whatever value of $g_0$ is chosen, it is always true that $g\to 0$ as $t\to \infty$. For SU(2)$_a$ 
we have $S_2(F)=3/2$, $S_2(S)=4$ (we have $4=2+2$ instead of $2$ because $A$ is complex)  and so the constant $b$ for the corresponding gauge coupling $g_a$ is
\be b_a=\frac{14}{3},\label{banoAp} \ee
which, being positive\footnote{One could extend the minimal model by adding other Weyl fermions charged under both SU(3)$_c$ and SU(2)$_a$ as long as the AF conditions remain satisfied. This would generically promote $y$ to a matrix.}, gives an AF $g_a$. 
With a similar computation one finds that the constant $b$ corresponding to SU(3)$_c$ is instead
\be b_s=\frac{29}{3}-\Delta,  \label{gsRGE}\ee
where $\Delta$ is the positive extra contribution due to the fermions and scalars in the SM sector. 
Using, for example, the results of~\cite{Pelaggi:2015kna} we find that  it is possible to have $g_s$ AF  keeping the  SM sector TAF. Moreover, since  $q$ and $\bar q$ do not have Yukawa couplings with the SM  sector (they transform under SU(2)$_a$ and U(1)$_{\rm PQ}$, but the fields in the SM sector do not), these extra quarks  favor the total asymptotic freedom in the SM sector: this is because the smaller $b_s$ is (keeping $b_s>0$) the bigger $g_s$ at a fixed energy favoring AF for the Yukawa and scalar quartic in the SM sector~\cite{Pelaggi:2015kna}.

\begin{figure}[t]
\begin{center}
\includegraphics[scale=0.6]{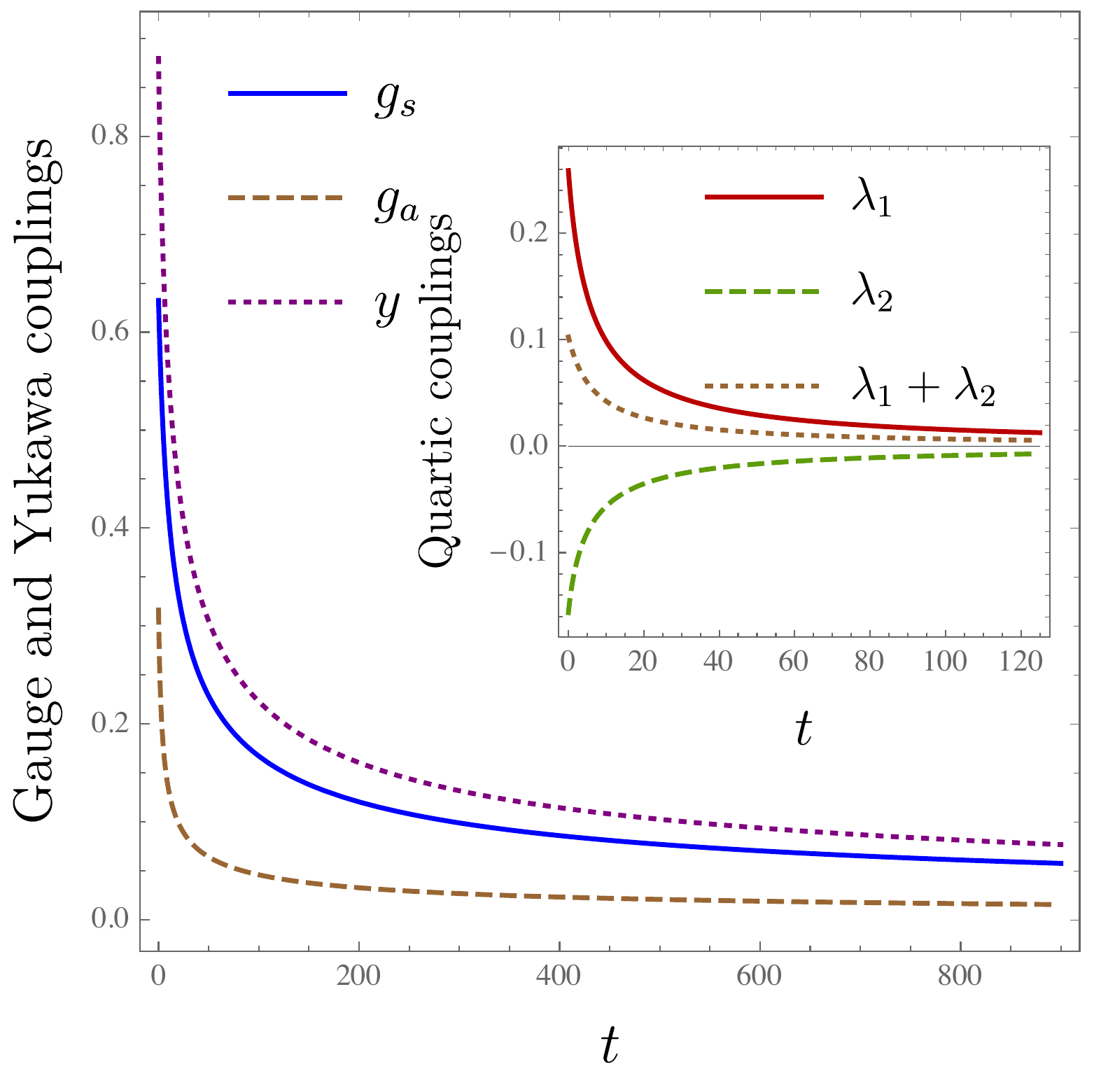} 
\end{center}
	\caption{\em  
	Running of couplings. They all flow to zero in the UV.
 In the plot we set $\Delta=28/3$ (compatibly with  known TAF SM sectors). The value $t$ $=$ $0$ is interpreted as the PQ symmetry breaking scale.}
\label{yff}
\end{figure}

\begin{figure}[t]
\begin{center}
\includegraphics[scale=0.55]{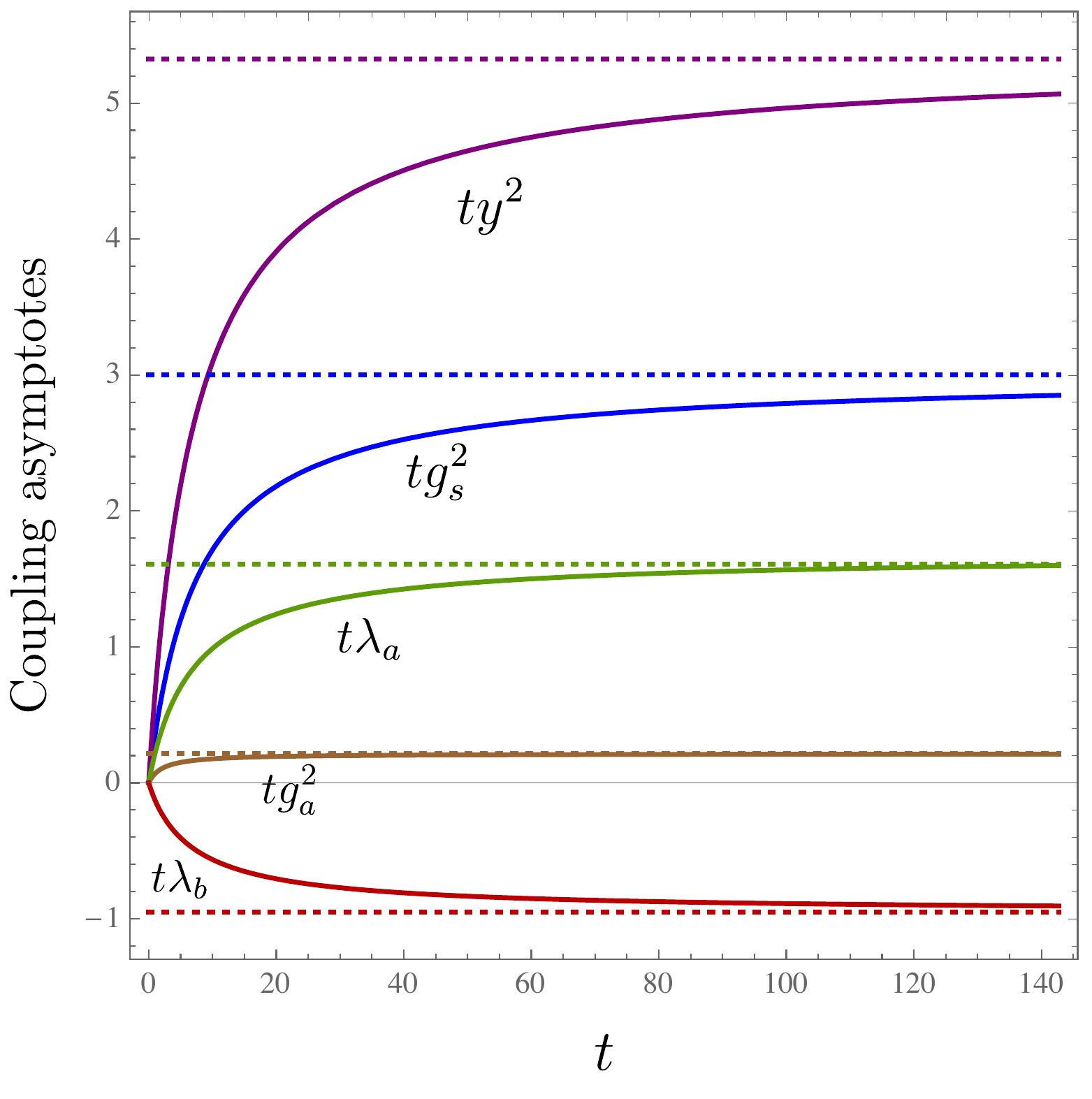} 
\end{center}
	\caption{\em  Solutions of the RGEs (multiplied by $t$) and their asymptotes as dictated by the fixed-flow ansatz. The parameters are set as in Fig.~\ref{yff}.}
\label{asymptotes}
\end{figure}

The RGE of $y$ is instead
\be\frac{dy^2}{dt} = y^2\left(\frac{9 y^2}{2}-8 g_s^2-\frac{9 g_a^2}{2} \right).\label{yRGE}\ee
Equations of this type have been studied in~\cite{Giudice:2014tma}. In our case the general solution to~(\ref{yRGE}) for any $b_a$ and $b_s$ is 
\be y^2(t) = \frac{y_0^2\left(1-\frac{9y_0^2I(t)}{2}\right)^{-1}}{\left(1+g_{s0}^2 b_st\right)^{8/b_s}\left(1+g_{a0}^2 b_at\right)^{9/(2b_a)}},\label{ySol} \ee
where $y_0\equiv y(0)$, $g_{s0}\equiv g_s(0)$, $g_{a0}\equiv g_a(0)$ and
\be I(t)\equiv \int_0^t  \frac{dt'}{\left(1+g_{s0}^2 b_st'\right)^{8/b_s}\left(1+g_{a0}^2 b_at'\right)^{9/(2b_a)}}.\label{Iexpr}\ee
We find that $I(t)$ admits a closed form expression\footnote{Indeed, the integral in~(\ref{Iexpr}) is a particular case of
\be \int_0^t  \frac{dt'}{\left(1+a_1t'\right)^{e_1}\left(1+a_2t'\right)^{e_2}}={\cal I}(t)-{\cal I}(0), \ee
where
\be {\cal I}(t) = \frac{  \left(\frac{a_1+a_1 a_2 t}{a_1-a_2}\right)^{e_2} \, _2F_1\left(1-e_1,e_2;2-e_1;\frac{-a_2-a_1 a_2 t}{a_1-a_2}\right)}{(1+a_1 t)^{e_1-1}(1+a_2 t)^{e_2}(a_1-a_1 e_1)},\ee
and $\, _2F_1$ is Gauss's hypergeometric function.}.
Looking at~(\ref{ySol}) and~(\ref{Iexpr}) and using the AF conditions for the gauge couplings ($b_s>0$ and $b_a>0$) we see that $y$ is AF if and only if $y_0$   satisfies 
\be y_0^2\leq \frac{2}{9I_\infty}, \qquad I_\infty\equiv\lim_{t\to\infty}I(t), \label{AFy} \ee
otherwise $y$ has a LP.
Note that if $b_s>0$ and $b_a>0$ the integral $I_\infty$ is positive and  convergent  whenever $8/b_s +9/(2b_a) >1$,
which, from~(\ref{banoAp}) and~(\ref{gsRGE}), is   satisfied for any value of $\Delta$ such that $b_s>0$, namely $\Delta<29/3$. This bound is compatible with the values in existing TAF SM sectors discussed in the literature. It follows that by taking $y_0$ small enough (such  that it  satisfies the  inequality in~(\ref{AFy})) one can indeed have an AF $y$. When the condition in~(\ref{AFy}) is fulfilled as a strict inequality $y$ decreases faster than the gauge couplings at large $t$. This class of solutions are UV attractive because if we perturb the initial condition $y_0$ by a small enough amount (keeping the inequality in~(\ref{AFy}) satisfied) the solution remains AF. 
When instead $y_0^2= 2/(9I_\infty)$ the Yukawa coupling decreases   like the gauge coupling at large $t$ (see Figs.~\ref{yff} and~\ref{asymptotes}).
Such solution   is not UV attractive, but  IR attractive because it requires a specific isolated value of $y_0$ (see below for a formal proof). This provides us with an interesting prediction of $y$ at low energy and, therefore, of the masses of the new quarks as discussed below in  Sec.~\ref{Stationary points and the mass spectrum}.

 The RGEs of $\lambda_1$ and $\lambda_2$ are $\frac{d\lambda_1}{dt}  = \beta_1$, and $\frac{d\lambda_2}{dt} =\beta_2$,
where
\be \beta_1(g, y, \lambda) =\frac92 g_a^4+\lambda_1 \left(8 \lambda_2+6 y^2-12 g_a^2\right)+14 \lambda_1^2+8 \lambda_2^2-3 y^4\ee
and 
 \be \beta_2(g, y, \lambda)  =\frac32 g_a^4+\lambda_2 \left(12 \lambda_1+6 y^2-12 g_a^2\right)+6 \lambda_2^2 +\frac32 y^4. 
 \ee
 
 The $\beta$-functions above have been obtained by applying the general formalism~of~\cite{Machacek:1983tz,Machacek:1983fi,Machacek:1984zw} to the present model.
The RGEs of  the $\lambda_i$ are too complicated for us to determine analytically the general solution at any $t$. However, we can understand if all couplings are AF by considering the ansatz
\be g_s^2(t) =\frac{\tilde g_s^2}{t}, \quad g_a^2(t) =\frac{\tilde g_a^2}{t}, \quad y^2(t) =\frac{\tilde y^2}{t}, \quad \lambda_i(t) =\frac{\tilde\lambda_i}{t}, \label{1overt} \ee
where $\tilde g_s^2$, $\tilde g_a^2$, $\tilde y$ and $\tilde\lambda_i$ are constants. The ansatz above  is manifestly TAF and is a fixed flow: although the couplings individually run, their ratios do not.
A solution of the form in~(\ref{1overt}) exists if and only if the corresponding algebraic system of equations obtained by plugging~(\ref{1overt}) into the RGEs admits solutions with $\tilde g_s^2$, $\tilde g_a^2$, $\tilde y^2$ and $\tilde\lambda_i$  real and also $\tilde g_s^2$,  $\tilde g_a^2$ and $\tilde y^2$ positive. Note that, when this condition is satisfied,~(\ref{1overt}) not only is a  solution of the RGEs, but also describes the $t\gg 1$ asymptotic behavior of any solution.

Let us first consider the  RGEs of the gauge couplings with the fixed-flow ansatz. Here we are interested in the case $\tilde g_s^2\neq 0$ (as we want to match the non-trivial low energy QCD running) and $\tilde g_a^2\neq 0$ because we want a TAF model. Then from~(\ref{betag}) $\tilde g_a^2 =1/b_a$ and  $\tilde g_s^2 =1/b_s$. 
Turning to the Yukawa coupling, we have either $\tilde y^2=0$ or 
\be \tilde y^2 = \frac{2}{9} \left(\frac{9}{2b_a}+\frac{8}{b_s}-1 \right).  \label{fixedf}\ee
The latter case corresponds to saturating the bound in~(\ref{AFy}) and is, therefore, an IR attractive solution as mentioned above. 
Finally the corresponding system of algebraic equations  for the quartic couplings reads
\be \tilde\lambda_i = -\beta_i(\tilde g,\tilde y, \tilde \lambda). \label{algebrlambda} \ee

In Table~\ref{suTAF} we show the real solutions $(\tilde\lambda_1,\tilde\lambda_2)$ to Eq.~(\ref{algebrlambda})  obtained by varying $\Delta$ (considering as an example the values corresponding to the TAF SM sector reported in~\cite{Pelaggi:2015kna}). In that Table $y$ is at the fixed-flow  in~(\ref{fixedf}). Taking instead the Yukawa coupling outside the fixed flow, that is setting $\tilde y=0$, produces  no TAF solutions.

\begin{table}[t]
\centering
\begin{tabular}{ |p{0.8cm} |p{3.1cm}|p{2.7cm}| }
\hline
\rowcolor{red!80!green!40!yellow!10}
$ \Delta $   & unstable vacuum & stable vacuum \\
\hline
28/3   & $(0.183, -3.23)$ & $(1.68, -0.951)$  \\
26/3    & $(0.149, -1.05)$ & $(0.575, -0.343)$\\
8    & $(0.145, -0.598)$ & $(0.349, -0.231)$\\
\hline
\end{tabular}
\caption{\it Real solutions $(\tilde\lambda_1,\tilde\lambda_2)$ to~(\ref{algebrlambda}) (corresponding to TAF solutions)  obtained by varying the contribution $\Delta$  to the RGE of the strong coupling (compatibly with a TAF SM sector, see e.g.~\cite{Pelaggi:2015kna}). The Yukawa coupling is at the fixed-flow, Eq.~(\ref{fixedf}).   The values of $(\tilde\lambda_1,\tilde\lambda_2)$ are approximated with three digits.}
\label{suTAF}
\end{table}

Note that the last column in Table~\ref{suTAF}   satisfies the vacuum stability condition in~(\ref{VacS12}), while the second column does not and the corresponding solutions are then ruled out. 
Furthermore, $\lambda_2$ is always negative. These features are quite robust and persist  even if we vary $b_a$ in addition to $b_s$. This can be done, for example, by adding a certain number $n_e$ of extra vector-like Dirac fermions, which are neutral under SU(3)$_c$ and U(1)$_{\rm PQ}$, but in the fundamental of SU(2)$_a$. The values of $(\tilde\lambda_1,\tilde\lambda_2)$ for all TAF solutions are then shown in Table~\ref{TAFne}. 

  We find that the solutions in the last column of  both Table~\ref{suTAF} and~\ref{TAFne} are all IR attractive, which results in a prediction for  the $\lambda_i$ at low energies and for the scalar spectrum, as discussed below in  Sec.~\ref{Stationary points and the mass spectrum}. 

In order to show that $y$ and $\lambda_i$ are IR attractive one can use the general formalism in~\cite{Giudice:2014tma}. According to this article, $y$ is IR attractive (repulsive) when the following quantity is positive (negative):
\be S_y \equiv \frac12 + \frac{\partial\beta_y}{\partial y}(\tilde g, \tilde y, \tilde \lambda), \ee 
where $\beta_y(g,y,\lambda)$ is the $\beta$-function of $y$ defined as $\beta_y \equiv dy/dt$. By using~(\ref
{banoAp}),~(\ref{gsRGE}) and~(\ref{fixedf}) we find that $S_y>0$ whenever the AF conditions for the gauge couplings ($b_s>0, b_a>0$) are satisfied.
Therefore, $y$ is IR attractive. Analogously, $\lambda_i$ is IR attractive (repulsive) when the following quantity is positive (negative)
\be S_i\equiv  1+\frac{\partial\beta_{\lambda_i}}{\partial {\lambda_i}}(\tilde g, \tilde y, \tilde \lambda).\ee
For all values in the last column of Table~\ref{suTAF} and~\ref{TAFne} we find $S_i>0$  so $\lambda_i$ are both  IR attractive when the unavoidable requirement of high-field stability is imposed.

\begin{table}[t]
\centering
\begin{tabular}{ |p{0.8cm} |p{0.3cm}|p{3.1cm}|p{2.7cm}| }
\hline
\rowcolor{red!80!green!40!yellow!10}
$ \Delta $ & $ n_e$ & unstable vacuum & stable vacuum \\
\hline
28/3  & 1&  $(0.219, -3.25)$   & $(1.70, -0.965)$ \\
{\tiny$//$} & 2  & $(0.268, -3.27)$ & $(1.73, -0.986)$ \\
{\tiny$//$} & 3  & $(0.344, -3.30)$ & $(1.77, -1.02)$\\
{\tiny$//$} & 4  & $(0.469, -3.34)$ & $(1.84, -1.08)$\\
{\tiny$//$} & 5  & $(0.722, -3.42)$ & $(1.97, -1.20)$\\
{\tiny$//$} & 6  & $(1.50, -3.49)$ & $(2.34, -1.70)$ \\
26/3 & 1  & $(0.185, -1.06)$ & $(0.593, -0.362)$\\
{\tiny$//$} & 2  & $(0.237, -1.07)$ & $(0.619, -0.389)$\\
{\tiny$//$} & 3  & $(0.314, -1.08)$ & $(0.656, -0.435)$\\
{\tiny$//$} & 4  & $(0.447, -1.08)$ & $(0.712, -0.528)$\\
8 & 1  & $(0.182, -0.601)$ & $(0.365, -0.255)$\\
{\tiny$//$} & 2  & $(0.236, -0.599)$ & $(0.387, -0.294)$\\
{\tiny$//$} & 3  & $(0.324, -0.570)$ & $(0.411, -0.376)$  \\
\hline
\end{tabular}
\caption{\it Real solutions $(\tilde\lambda_1,\tilde\lambda_2)$ as in Table~\ref{suTAF} except that $n_e$ vector-like Dirac fermions (in the fundamental of SU(2)$_a$, but neutral under SU(3)$_c$ and U(1)$_{\rm PQ}$)  are added. The number $n_e$ is varied until total asymptotic freedom is possible. }
\label{TAFne}
\end{table}

We can also find numerically the solutions to the RGEs of the quartic couplings for {\it any} given initial conditions of the gauge couplings even outside the fixed-flow ansatz  in~(\ref{1overt}). In Fig.~\ref{yff} we also plot the running of the quartic couplings and their sum (to show, among other things, that the high-field stability conditions in~(\ref{VacS12}) are satisfied without relying on the fixed-flow ansatz). Choosing perturbative low-energy values of the gauge couplings results in predicted perturbative values of the Yukawa and quartic couplings as shown in Fig.~\ref{yff}. Therefore, our one-loop approximation for the predicted values is reliable. In Fig.~\ref{asymptotes} it is shown that also the quartic couplings  (like the Yukawa one) scale as the gauge couplings in the $t\to\infty$ limit: indeed, $\lambda_i$ approach $\tilde\lambda_i$ as dictated by Eq.~(\ref{1overt}). However, at low energies the running goes generically outside the fixed-flow ansatz as clear from  Fig.~\ref{asymptotes}.

\section{Stationary points and the mass spectrum}\label{Stationary points and the mass spectrum}
The two Hermitian adjoint representations $A_R$ and $A_I$  can be expressed in terms of the Pauli matrices $\sigma^k$ as follows: $A_R = A_{Rk}\sigma^k/2$,  $A_I =   A_{Ik}\sigma^k/2,$ where a sum over $k=1,2,3$ is understood.  
  Since ${\rm SU(2)}_a$ transforms the $A_{Rk}$ and $A_{Ik}$ as ordinary rotations transform the  coordinates in three dimensions, it is possible to set $A_{R3}=A_{I2}=A_{I3}=0$ through an ${\rm SU(2)}_a$ transformation. In the following we, therefore, do so without loss of generality. 
  
  For general values of $A_{R1}$, $A_{I1}$ and $A_{R2}$ the three SU(2)$_a$ gauge fields acquire  the following  masses:
\be   M_V = g_a\sqrt{A_{R1}^2+A_{R2}^2+A_{I1}^2}, \qquad M_{V\pm}=\sqrt{\frac{M_V^2}{2}\pm \sqrt{\frac{M_V^4}{4}-g_a^4A_{I1}^2A_{R2}^2}}.\label{VecMass}\ee
 The Weyl quarks $q$ and $\bar q$ (that are doublets under ${\rm SU(2)}_a$) form instead two Dirac quarks $Q_{\pm}$ (both triplets under ${\rm SU(3)}_c$) with masses
    \be M_{Q\pm} = \frac{y}2 \sqrt{A_{R1}^2+(A_{I1}\pm A_{R2})^2}. \label{MQtilde}\ee

There are only three physically inequivalent stationary points. 
An obvious one is  the origin ($A_{R1}=A_{I1}=A_{R2}=0$), which leaves SU(2)$_a$ unbroken and corresponds to a maximun of $V_A$. Next, there are all the equivalent configurations obtained from
  \be A_{R1}= \frac{m}{\sqrt{\lambda_1+\lambda_2}}, \qquad A_{I1} =0, \qquad A_{R2}=0, \label{VacRes} \ee
  through a U(1)$_{\rm PQ}$ and/or an SU(2)$_a$ transformation.  Note that $A_{R1}$ in~(\ref{VacRes}) is guaranteed to be real from the high-field stability condition in~(\ref{VacS12}). These stationary points break SU(2)$_a$ down to a residual Abelian group U(1)$_a$. Indeed,  from~(\ref{VecMass}) one has
$M_{V-} = 0$ , and $M_{V+} = M_V =g_a m/\sqrt{\lambda_1+\lambda_2}$.  
  The value of $A_{R1}$ given by~(\ref{VacRes}) is the PQ symmetry breaking scale $f_a$.
    The corresponding value of $V_A$ is
  \be V_A = -\frac{m^4}{4(\lambda_1+\lambda_2)}\qquad (\mbox{potential at~(\ref{VacRes})}). \label{VAres}\ee
The quarks $Q_{\pm}$ acquire equal non-vanishing masses for $y\neq 0$:
 $M_{Q+}=M_{Q-}= y \, m/(2\sqrt{\lambda_1+\lambda_2})$.
The scalar spectrum corresponding to~(\ref{VacRes}) includes  three massive scalars with squared masses
$M_{S1}^2=2m^2,$  and
$M_{S2}^2=M_{S3}^2=  -2\lambda_2 m^2/(\lambda_1+\lambda_2)$. 
Note that also the second squared mass is positive for $\lambda_2<0$ and $\lambda_1+\lambda_2>0$, which is the case for the TAF solutions with stable vacuum reported in the last column of Tables~\ref{suTAF} and~\ref{TAFne}. The vacuum in~(\ref{VacRes}) is, therefore, a minimum of the potential (and actually, as we will see, the absolute minimum) using the TAF and high-field stability requirements. 
  The scalar spectrum   also includes three massless modes, two of them are eaten by the two massive vector bosons. The third one is the axion. Since ${\rm U(1)}_{\rm PQ}$ is broken by anomalies the axion receives as usual a mass at quantum level.

   The last class of stationary points consists of all the equivalent configurations obtained from 
   \be A_{R1} = 0, \qquad A_{I1}=\pm A_{R2}=\pm  \frac{m}{\sqrt{2\lambda_1}} \label{VacBreak} \ee
     through a U(1)$_{\rm PQ}$ and/or an SU(2)$_a$ transformation. These configurations break SU(2)$_a$  completely; indeed, from~(\ref{VecMass}) one finds $M_V =\sqrt{2}g_a|A_{I1}|$ and $M_{V\pm} =g_a |A_{I1}|$, but are not phenomenologically acceptable because they lead to a massless extra colored fermion (see~(\ref{MQtilde})).  
     Inserting~(\ref{VacBreak}) into $V_A$ one obtains $V_A = -m^4/(4\lambda_1)$. Since $\lambda_1$ must be positive from  high-field stability this value of $V_A$ is higher than the one in~(\ref{VAres}) if and only if $\lambda_2<0$ and $\lambda_1+\lambda_2>0$, which is the case for the TAF solutions reported in the last column of Table~\ref{suTAF} and~\ref{TAFne} (those with a stable vacuum). Moreover, the scalar squared mass matrix  in case~(\ref{VacBreak}) has eigenvalues $2m^2\lambda_2/\lambda_1$ and $2m^2$. So  the TAF requirement and high-field stability automatically allow (and actually force) us to exclude the phenomenologically unacceptable stationary points in~(\ref{VacBreak}) because they guarantee that the vacuum in~(\ref{VacRes}) is the {\it absolute} minimum of the potential and the stationary points in~(\ref{VacBreak}) are only  saddle points.

Note that having a vacuum with the residual U(1)$_a$ is not phenomenologically ruled out. Indeed, one can perform a linear homogeneous transformation on the ordinary hypercharge and the U(1)$_a$ gauge bosons in a way that the extra massless boson (which appears at low energies as a dark photon) does not interact at the renormalizable  level with the SM particles; its effective interactions can be generated only  via loop contributions involving the extra quarks $q$ and $\bar q$. As long as the masses of these fermions, $M_{Q\pm}$, are large enough these interactions appear at low energies as non-renormalizable terms in the Lagrangian suppressed by appropriate powers of the large masses.  The dark photon is compatible with the observations given that   $M_{Q\pm}$ are around the PQ symmetry breaking scale $f_a$. Indeed, $f_a$ is at least of order $10^8$~GeV and even higher to account for the whole dark matter through the axion (see~\cite{DiLuzio:2020wdo} for a recent review on axion bounds), which is more than enough to satisfy the observational bounds~\cite{Dobrescu:2004wz}. 

For example, let us consider the limit on the number of effective relativistic degrees of freedom (see~\cite{Fields:2019pfx} for a recent determination), which implies the dark photon decouples at a temperature $T_d >T_{\rm BBN}\sim$ MeV. As shown in~\cite{Dobrescu:2004wz}, this  translates into a bound on the mass scale 
 suppressing the effective interaction between the dark photon and the Higgs boson, the quarks and the leptons of the SM. The exact form of this bound depends on the precise way the dark photon can interact via loops with the SM. But in any case the bound found in~\cite{Dobrescu:2004wz} is not exceeding more than two orders of magnitude the TeV scale and so is amply satisfied by the dark photon of the TAF axion sector, whose interactions with the SM fields are suppressed, as discussed above, by a scale at least as large as $10^8~$GeV. We also observe that the dark photon does not produce modifications on the spectrum of isocurvature perturbations exceeding the observational bounds (the most recent ones are those by Planck~\cite{Ade:2015lrj}). Indeed,  the dark photon, being a massless spin-1 particle, contributes only with  vector modes to the cosmological perturbations and those modes are known to decay with time.

Finally, we note that the requirement of TAF couplings and vacuum stability leads to a prediction for the scalar masses $M_{S1}$, $M_{S2}$ and $M_{S3}$   and for the masses of the new quarks, $M_{Q\pm}$. This is because $y$ and $\lambda_i$ are predicted at low energies by the TAF requirement once $g_a$ and $g_s$ are fixed at low energies and, therefore,  the above-mentioned masses can be extracted once the mass of the extra massive spin-1 particle is fixed. To make this explicit note that 
\be M_{S1}=\sqrt{2} f_a\sqrt{\lambda_1(t_a)+\lambda_2(t_a)}, \qquad M_{S2}=M_{S3}=\sqrt{-2\lambda_2(t_a)} f_a,\qquad  M_{Q\pm} = y(t_a)f_a/2, \ee 
where $t_a$ is the low energy value of $t$ (namely $t$ computed at the PQ scale) and the PQ symmetry breaking scale can be written as follows, $f_a = M_V/g_a(t_a)$. One can set, for example, $t_a=0$ without loss of generality by choosing appropriately the arbitrary reference scale $\mu_0$ (as done in Figs.~\ref{yff} and~\ref{asymptotes}). 
Our result here is opposed to known (non-TAF) axion models, where the masses and couplings of the new particles are freely adjustable parameters. The fact that the Yukawa and quartic couplings as well as the  masses $M_{S1}$, $M_{S2}$, $M_{S3}$ and $M_{Q\pm}$ are  predicted at low energies can lead to testable predictions for cosmology. One way one could test this model is through gravitational wave detectors; the spectrum of gravitational waves produced by the PQ symmetry breaking has specific features due to the fact that the theory has less adjustable parameters than in non-TAF axion models~\cite{Anish}.
    
    \section{Conclusions}

  A fundamental field theory of the QCD axion has to have certain features. In particular, the axion sector should be invariant under a non-Abelian gauge group to ensure total asymptotic freedom. Here, the minimal realistic model of this sort has been explicitly built and studied: it features an SU(2)$_a$ gauge symmetry, a complex scalar $A$ in the adjoint representation of SU(2)$_a$ and one extra Dirac field $\{q,\bar q\}$ in the fundamental representation of SU(3)$_c\times$SU(2)$_a$ to implement the U(1)$_{\rm PQ}$ symmetry. All PQ charges of the SM particles have been set to zero for simplicity. We have shown that there are initial conditions for the RG flow such that the model is TAF and features an absolutely stable vacuum at the same time. An interesting feature of this  model is the presence of a dark photon in the low-energy spectrum.
    
 Besides the presence of extra non-Abelian gauge symmetries a generic TAF theory can predict a number of observable quantities given that the RG flow typically involves IR attractive fixed points. In the minimal model proposed, indeed, we have seen that some of the masses of the extra particles are predicted in terms of other  parameters that  would have been independent in an effective model with   a  finite cutoff. This is the case for the extra quarks and scalars,  whose masses can  be expressed in terms of the SU(3)$_c$ and  SU(2)$_a$ gauge couplings and the SU(2)$_a$ vector boson mass or, equivalently, $f_a$. The reason is that the corresponding Yukawa and quartic couplings are IR attractive to realize the TAF requirement.
     
       Let us conclude by giving some examples of possible outlook. It would be interesting to construct  TAF models of the QCD axion where the quarks carrying the PQ charges are those already present in the SM. For example, one could construct a  DFSZ-like~\cite{DFSZ}  TAF model. This could have interesting implications for the Higgs physics given that the DFSZ model features an extra Higgs doublet. 
        Also, it would be valuable to know whether the dark photon present in the low energy spectrum of the minimal model generically appears in other   TAF axion models. Another example of possible outlook is the construction of fundamental QCD axion models where some couplings flow to an interacting  UV fixed point.

    \vspace{0.4cm}
    
\subsection*{Acknowledgments}
I thank R.~Frezzotti and A.~Ghoshal  for interesting discussions.

\vspace{0.5cm}


\end{document}